\documentclass[preprint,preprintnumbers,amsmath,amssymb,nofootinbib,toc,12pt]{revtex4}
\usepackage{latexsym}
\usepackage{graphicx}
\usepackage{amsthm}
\usepackage{float}
\usepackage{amsmath}
\usepackage{graphicx,subfigure}
\usepackage{fancybox,feynmp}
\usepackage{indentfirst}
\usepackage{epsfig}
\usepackage{epsfig,subfigure,psfrag}
\usepackage{dcolumn}
\usepackage{bm}
\usepackage{slashed}
\usepackage{cancel}
\usepackage{color}
\usepackage{tabularx}
\usepackage{hyperref}
\hypersetup{
    colorlinks=true,       
    linkcolor=blue,          
    citecolor=blue,        
    filecolor=blue,      
    urlcolor=blue           
}

\preprint{CERN-PH-TH-2015-024}

\newcommand{\be}{\begin{eqnarray}}
\newcommand{\ee}{\end{eqnarray}}
\newcommand{\beq}{\begin{eqnarray}}
\newcommand{\eeq}{\end{eqnarray}}
\newcommand{\lag}{\mathcal L}
\newcommand{\bea}{\begin{eqnarray}}
\newcommand{\eea}{\end{eqnarray}}

\newcommand{\GeV}{{~\rm GeV}}
\newcommand{\gev}{{~\rm GeV}}

\newcommand{\tev}{{~\rm TeV}}

\newcommand{\vev}[1]{{\langle #1 \rangle}}

\newcommand{\da}[1]{{\bf \color{magenta} DA: #1}}

\newcommand{\mb}[1]{\mathbf{#1}}
\def\Tr{\mathop{\rm Tr}}
\newcommand{\Um}{\text{U}}
\newcommand{\SU}{\text{SU}}

\newcommand{\Eq}[1]{Eq.~(\ref{#1})}

\begin{document}

\title{Models of Goldstone Gauginos}

\author{Daniele S. M. Alves}
\email{spier@nyu.edu}
\affiliation{Center for Cosmology and Particle Physics, Department of Physics, New York University, New York, NY 10003}
\affiliation{Department of Physics, Princeton University, Princeton, NJ 08544}

\author{Jamison Galloway}
\email{jamison.galloway@nyu.edu}
\affiliation{Center for Cosmology and Particle Physics, Department of Physics, New York University, New York, NY 10003}

\author{Matthew McCullough}
\email{matthew.mccullough@cern.ch}
\affiliation{Theory Division, CERN, 1211 Geneva 23, Switzerland}

\author{Neal Weiner}
\email{neal.weiner@nyu.edu}
\affiliation{Center for Cosmology and Particle Physics, Department of Physics, New York University, New York, NY 10003}

\date{\today \\ \vspace*{1cm}}

\begin{abstract}
Models with Dirac gauginos provide appealing scenarios for physics beyond the standard model. They have smaller radiative corrections to the Higgs mass, a suppression of certain SUSY production processes, and ameliorated flavor constraints. Unfortunately, they also generally have tachyons, the solutions to which typically spoil these positive features. The recently proposed ``Goldstone Gaugino'' mechanism provides a simple solution that eliminates these tachyonic states. We provide details on this mechanism and explore models for its origin. In particular, we find SUSY QCD models that realize this idea simply, and discuss scenarios for unification. 
\end{abstract}

\maketitle

\tableofcontents

\newpage

\numberwithin{equation}{section}
\renewcommand\theequation{\arabic{section}.\arabic{equation}}
\section{\label{sec:intro} Introduction}
The turn on of the LHC has challenged a number of basic ideas about naturalness. While parameter regions for supersymmetry and other solutions to the hierarchy problem remain, large swaths of ``natural'' parameter space have been eliminated, causing us to rethink whether the weak scale is, itself, natural \cite{Wells:2003tf,ArkaniHamed:2004fb,Giudice:2004tc,Fox:2005yp,Arvanitaki:2012ps,ArkaniHamed:2012gw}.

A great challenge for a number of these scenarios comes from the renormalization group flow. In the MSSM, for instance, large stop masses feed into the Higgs soft masses quickly, meaning that the stop masses should be ideally generated at a low scale. At the same time, existing searches for gluinos have been pushing their masses up, which, in turn, exacerbate their corrections to the stops and Higgs soft masses. 

A simple resolution of this could lie in a departure from the framework of the MSSM. One simple deviation is to break supersymmetry with Dirac gauginos \cite{Fayet:1978qc,Polchinski:1982an,Girardello:1997hf,Polonsky:2000zt,Hall:1990hq,Randall:1992cq,Nelson:2002ca,Fox:2002bu}, rather than Majorana. A remarkable effect of the presence of the Dirac gaugino mass is that it is ``supersoft'' \cite{Fox:2002bu} and contributes only a finite threshold to the soft scalar masses. As a consequence, there is no sizable running contributing to the Higgs mass, and even very heavy squarks (stops) would introduce only mild $\sim 1\-- 10\%$ tuning \cite{Fox:2002bu}.

Moreover, this supersoftness can result in dramatic phenomenological effects. The leading effect is to have gauginos which are naturally 5-10 times heavier than the scalar sector, which is at odds with conventional expectations from the MSSM. By allowing this, a number of other possibilities arise. First, because the gauginos are both heavy and Dirac, the production cross section for colored scalars is suppressed, weakening limits on them \cite{Kribs:2012gx}. Heavy Dirac gauginos also weaken flavor and CP constraints considerably \cite{Kribs:2007ac}, alleviating the SUSY flavor and CP problems. These setups can naturally arise from a number of UV scenarios \cite{Fayet:1978qc,Polchinski:1982an,Girardello:1997hf,Polonsky:2000zt,Hall:1990hq,Randall:1992cq,Nelson:2002ca,Fox:2002bu,Antoniadis:2005em,Antoniadis:2006eb,Antoniadis:2006uj,Kribs:2007ac,Amigo:2008rc,Blechman:2009if,Plehn:2008ae,Benakli:2008pg,Choi:2008ub,Belanger:2009wf,Benakli:2009mk,Choi:2009ue,Benakli:2010gi,Choi:2010gc,Carpenter:2010as,Kribs:2010md,Abel:2011dc,Davies:2011mp,Benakli:2011kz,Kalinowski:2011zzc,Frugiuele:2011mh,Itoyama:2011zi,Itoyama:2013sn,Itoyama:2013vxa,Rehermann:2011ax,Bertuzzo:2012su,Argurio:2012cd,Goodsell:2012fm,Fok:2012fb,Argurio:2012bi,Frugiuele:2012pe,Frugiuele:2012kp,Benakli:2012cy,Kribs:2012gx,Kribs:2013oda,Kribs:2013eua,Chakraborty:2013gea,Csaki:2013fla,Beauchesne:2014pra,Bertuzzo:2014bwa,Benakli:2014cia,Dimopoulos:2014aua,Chakraborty:2014tma}.

From this, it would seem that Dirac gauginos are a promising approach to solving the naturalness issues of supersymmetry. There are, however, a few problems \cite{Fox:2002bu}: the existence of the supersoft operator naturally is accompanied by a $B\mu$-type term, which induces a tachyon for either the scalar or pseudo-scalar in the adjoint; RG running tends to spread the gauginos widely, leading to very heavy gluinos; and finally, the $D$-term quartic for the Higgs boson is naturally suppressed.\footnote{See \cite{Csaki:2013fla} for a detailed discussion.} A singlet can ameliorate the lack of $D$-term quartics, and modifying assumptions about gaugino universality can compensate for the running, which pulls up the gluino. The $B\mu$ term, however, is quite robust, appearing ubiquitously in models that generate the Dirac gaugino mass. Consequently, any scenario that truly allows natural models of Dirac gauginos must first address this $B\mu$ problem. Addressing this would eliminate the last vestiges of objections to the Dirac gaugino scenario.

In \cite{Alves:2015kia}  a solution to the $B\mu$ problem of Dirac gauginos was proposed, coined the Goldstone Gaugino (GoGa) mechanism. There, the right-handed gaugino originates from a (pseudo)-Goldstone superfield of a spontaneously broken anomalous global symmetry. In the GoGa scenario, the dangerous operators are naturally absent, while the Dirac gaugino mass is still present and originates from a Wess-Zumino-Witten (WZW) term.  In this paper, we outline the generic features of the Goldstone Gaugino mechanism (Sec.~\ref{sec:GG}), and present natural UV completions to the GoGa scenario, one strongly-coupled in the context of SQCD with $F=N$ flavors (Sec.~\ref{sec:nonperturb}), and another one perturbative (Sec.~\ref{sec:perturb}). Finally, we discuss their implications for naturalness (Sec.~\ref{conc1}), gauge coupling unification (Sec.~\ref{conc2}), and collider phenomenology (Sec.~\ref{conc3}). We conclude in Sec.~\ref{concl}.


\section{\label{sec:GG} Goldstone Gauginos}
The Goldstone Gaugino mechanism is a realization of Dirac gauginos and has been described in \cite{Alves:2015kia}. Here we review the basics of Dirac gauginos as well as the GoGa mechanism, outlining its essential ingredients and how it addresses the $B\mu$ problem.

The Dirac gaugino mass marries the fermion in an adjoint chiral superfield $A$ with the fermion in the vector supermultiplet via the operator
\beq
\label{eq:classic}
\Delta W = \frac{\sqrt 2}{M_{\rm C}}W'^\alpha W^i_\alpha A^i,
\eeq
where $W'_\alpha$ is a spurion superfield for a $\Um(1)$ that acquires a $D$-term expectation value $\vev{W'_\alpha}=\theta_\alpha D$. 
This operator generates a Dirac mass $m_D=D/M_{\rm C}$ for the fermion, and a mass for the real scalar $m_{a_R} = 2 m_D$, while leaving the pseudoscalar massless. Note that this operator respects a shift symmetry under which the pseudoscalar shifts by a constant, $a_I \rightarrow a_I + \chi$.

The dangerous $B\mu$-like term arises from
\beq
\label{eq:lemontwist}
\Delta W = \frac{1}{2M_{\rm LT}^2}W'^\alpha W'_{\alpha} A^i A^i.
\eeq
This term\footnote{
Such a contribution is often referred to as the ``lemon-twist'' supersoft operator \cite{Carpenter:2010as}, distinguishing it from the ``classic'' supersoft operator of Eq.~(\ref{eq:classic}), hence the subscripts of the respective scales.  We will adopt this nomenclature here and refer to these operators as LTS and CS respectively. } 
generates a mass squared with opposite sign for $a_R$ and $a_I$, i.e. $\delta m_{a_R}^2 = -\delta m_{a_I}^2 = D^2/M_{\rm LT}^2$.  In principle, the theory as written need not be problematic. If the lemon-twist operator gives a negative mass squared to $a_R$, then in combination with the classic operator it may yield no tachyons.

As a matter of practical course, however, this is not the case. 
If both Eqs.~(\ref{eq:classic}, \ref{eq:lemontwist}) arise at the same loop order, then we naturally expect $m_{a_I} \sim 4 \pi m_D$, and the classic operator cannot cancel the tachyon from lemon-twist. This is aggravated by the fact that the classic operator comes with a $g_i$ for the gauge group indexed by $i$, which further suppresses it relative to lemon-twist.  A solution involving large additional $F$-term scalar masses for the adjoint breaks supsersoftness, and so the SM superpartners obtain masses fixed by a UV-sensitive logarithm rather than just the finite threshold, leading back to the need for tuning as in the conventional MSSM.  A superpotential mass $\Delta W = m_A A^2$, as the other simple alternative, is unappealing since its required size would render the SM gauginos almost purely Majorana \cite{Fox:2002bu,Arvanitaki:2013yja}.

The essential lemon-twist issue is easily seen in the context of the simplest scenario to generate Dirac gaugino masses, with messengers ($T, \overline T$) charged under the Standard Model and the $\Um (1)$ that carries the $D$-term VEV. With the superpotential
\be
W = \lambda \overline T A T  + \mu  \overline T T
\label{eq:messengermodel}
\ee
and the masses for the messenger scalars split by the $D$-term, $\mu^2 \pm D$,  one-loop diagrams generate both classic and lemon twist operators with coefficients 
\be
\frac{1}{M_{\rm C}} = \sqrt 2  \frac{g_i  \lambda }{16 \pi^2}\frac{1}{\mu}; \quad
\frac{1}{M_{\rm LT}^2} = -\frac{\lambda^2}{16 \pi^2} \frac{1}{\mu^2},
\label{eq:coefficients}
\ee
where $g_i$ is the appropriate SM gauge coupling. The problem seems intractable: since we may write both $W'^\alpha W^i_\alpha A^i$ and $W^{i\,\alpha} W^i_\alpha $,  no  obvious symmetry is able forbid the lemon-twist operator.

Ideally, we would impose a symmetry $a_I \rightarrow a_I+\chi$ with $\chi = {\rm constant}$, which would forbid lemon-twist but allow the classic operator. This breaks supersymmetry, however, and while a useful symmetry to keep in mind, it is difficult to implement.\footnote{Interestingly, this symmetry would allow an operator in the K\"ahler potential $\int d^4 \theta D_\alpha V' W_\alpha (A + A^\dagger)$ \cite{Csaki:2013fla} (``lime-twist''), which is not supersoft and is problematic in inducing large two-loop contributions to the soft masses of the MSSM sparticles. This operator is not gauge invariant, however, but could plausibly still arise if $V'$ is simply a spurion field, not representing any genuine gauge symmetry. Still, if the symmetry represented by $V'$ {\em could} have been gauged (in the sense that anomalies cancel and no other gauge non-invariant terms are present), which is the usual case, it will not arise radiatively. }
If we elevate the shift parameter $\chi$ to a chiral superfield and impose a super-shift symmetry $A \rightarrow A + \chi$, this necessarily forbids both the classic and lemon-twist terms. 

There is, however, a simple solution: namely, the Goldstone Gaugino mechanism \cite{Alves:2015kia}, whose essential ingredients we shall now review.

In the GoGa mechanism, a shift symmetry forbidding the lemon-twist operator emerges from the spontaneous breaking of an approximate global symmetry, which is explicitly violated by the SM gauge couplings. Moreover, the currents associated with the broken generators of the global symmetry have a mixed anomaly under the SM and the $\Um (1)$ that carries the $D$-term VEV. This mixed anomaly turns out to be the source of the Classic supersoft operator, identified with a WZW term \cite{gauginoWZW}. In this realization, the right handed gauginos originate from the Goldstones of the broken symmetry. In the limit $g\rightarrow 0$ the anomaly vanishes and so does the Classic supersoft operator, consistent with the restoration of an exact shift symmetry.

A simple toy model illustrating this mechanism is the following. Consider as before the messenger sector $T,\overline T$, now coupling to a bifundamental $\Sigma$ of an enlarged $\SU(F)_L\times \SU(F)_R\times\Um (1)_T$ global symmetry:
\be
\label{GGtoy}
W=\lambda\; \overline T\, \Sigma\, T. 
\ee
The Standard Model is embedded in the vector subgroup $\SU(F)_V$, and we assume the UV dynamics induces a VEV for $\Sigma$, $\langle \Sigma\rangle = v \times \bold{1}$, spontaneously breaking the global symmetry to $\SU(F)_V\times\Um (1)_T$. The surviving degrees of freedom are the Goldstone fields $\Pi$ associated with the broken generators, parameterized as
\be
\Sigma = (v+\sigma)\, e^{\Pi/f}.
\ee
The axial currents associated with the Goldstones $\Pi$ have a mixed $\SU(F)_A$-$\SU(F)_V$-$\Um (1)_T$ anomaly, which is realized in the IR by a (supersymmetric) WZW term:
\be
W_{\text{WZW}} \supset \frac{g}{32\pi^2f}\;W_T^\alpha\, W^i_\alpha\, \Pi^i \, . 
\ee
This is precisely the Classic supersoft operator, with the $D$-term of $W_T^\alpha$ identified as the SUSY breaking spurion.

The absence of the lemon-twist operator in the low energy theory can be demonstrated in several ways, which we do in the Appendix. In Appx.~\ref{sec:AToyModel} we show it diagrammatically by keeping track of the contributions to the quartic couplings from up to $\mathcal{O}(\Pi^2)$ terms in the UV superpotential. In Appx.~\ref{sec:hgc} we argue that the lemon-twist operator cannot be generated because the Goldstones $\Pi$, when treated as background fields, do not contribute to the messenger masses, and hence cannot correct the $\Um (1)_T$ holomorphic gauge coupling below the messenger threshold. Finally, in Appx.~\ref{appx:tadpole} we demonstrate the absence of lemon-twist in the linear realization of Goldstone gauginos, where a tadpole-induced shift in the VEV of the trace component of $\Sigma$ cancels the $B\mu$-terms that lead to tachyonic scalars.


\section{\label{sec:nonperturb} A Strongly-Coupled Model}

\subsection{Warm-up: non-renormalizable toy model}
\label{sec:NRtoy}

A crucial aspect of the GoGa mechanism that was not addressed in the previous discussion is the dynamics that induces the spontaneous breaking of the global symmetry.
This dynamics can be parameterized in the simple toy model (\ref{GGtoy}) by the inclusion of a non-renormalizable operator such as
\be
\label{GGtoy}
W=\lambda\; \overline T\, \Sigma\, T + \frac{1}{\Lambda^{F-2}}\,S\,\big(\text{Det}\Sigma - v^F\big),
\ee
where $S$ is a singlet whose $F$-term induces the desired VEV for $\Sigma$.

It turns out that this symmetry breaking parameterization automatically arises in SUSY QCD (SQCD) with $F=N$ flavors via the strong dynamics. In fact, all the basic ingredients for the Goldstone Gaugino mechanism are naturally present in this SQCD scenario: (i) the Goldstone fields arise as composite states of the microscopic quarks, with the composite meson $M=Q\overline Q$ playing the analogous role of $\Sigma$ in our toy model; (ii) the UV theory has the appropriate global symmetry, namely, $\SU(F)_L\times \SU(F)_R\times\Um (1)_B$ (iii) which is spontaneously broken via the strong dynamics in the desired pattern due to the quantum deformation of the moduli space and (iv) a WZW term is present in the IR to realize the mixed anomaly that sources the Classic supersoft operator. In order to fully implement this strongly-coupled model of Goldstone Gauginos, all that is left to do is to embed the Standard Model gauge symmetries in the diagonal flavor group, and induce a SUSY breaking $D$-term associated with the $U(1)_B$ baryon number.

This strongly-coupled model provides a coherent and elegant realization of the GoGa mechanism. In the following we will review some important aspects of $F=N$ SQCD, and discuss the GoGa implementation in detail. In the last subsection we will derive the Classic supersoft operator starting from $F=N+1$ SQCD and integrating out the flavored baryons to recover the $F=N$ limit.

\subsection{Composite Goldstones and SUSY QCD}
\label{sec:SQCD}
A minimal realization of the Goldstone gaugino scenario can arise in models of SUSY QCD that become strongly coupled in the IR once we weakly gauge a subgroup of the theory's flavor symmetry, and identify it with the SM.  These theories are described in the UV by an $\SU(N)$ gauge group with $F$ flavors of quarks denoted by $Q$ and $\overline Q$. The IR degrees of freedom are weakly coupled composite mesons $M=Q\overline Q$, and for $F\geq N$ also baryons $B=Q^N$ and $\overline B=\overline Q^N$.  The strong coupling scale is related to the lowest component of the holomorphic scale, $\Lambda^b$, with $b=3N-F$ the one-loop beta function for the $\SU(N)$ gauge coupling.
For reference, we list the symmetries and field content of this class of theories in table~\ref{tab:Gcharges}.
\begin{table}[ht]
\begin{center}
\renewcommand{\arraystretch}{1.}
\begin{tabular}{c | c | c c c c | c c}
 & $\SU(N)$ & $\SU(F)_L$ & $\SU(F)_R$ & $\Um(1)_B$ & $\Um(1)_A$ & $\SU(F)_V$ & $\Um(1)_{R}$ \\
\hline
$Q$ & $\square$ & $\square$ & $1$ & $1$ & $1$ & $\square$ & $\frac{F-N}{F} $ \\
$\overline Q$  & $\overline \square$ & $1$ & $\overline \square$ & $-1$ & $1$ & $\overline \square$ & $\frac{F-N}{F}$\\
\hline 
$M$ & $1$ & $\square$ & $\overline \square$ & $0$ & $2$ & $   1 + {\rm Adj.}$ & $2 \cdot \frac{F-N}{F}$ \\
$B$ & $1$ & $\overline{\binom{F}{N}}$ & $1$ & $N$ & $N$ & $\overline{\binom{F}{N}}$ & $N \cdot \frac{F-N}{F}$\\
$\overline B$ & $1$ & $1$ & $\binom{F}{N}$ & $-N$ & $N$ & $\binom{F}{N}$ & $N \cdot \frac{F-N}{F}$\\
\hline 
$\Lambda^b$ & $1$ & $1$ & $1$ & $0$ & $2F$ & $1$ & $0$ \\
\hline
\end{tabular}
\linespread{1}
\caption{\small{
Global symmetries and charges for the UV and IR degrees of freedom of SUSY QCD.  The diagonal flavor subgroup $\SU(F)_V$ is gauged and identified with the SM.}}
\label{tab:Gcharges}
\end{center}
\end{table} 

\begin{figure}[h]
\begin{center}
\includegraphics[width=6.5cm]{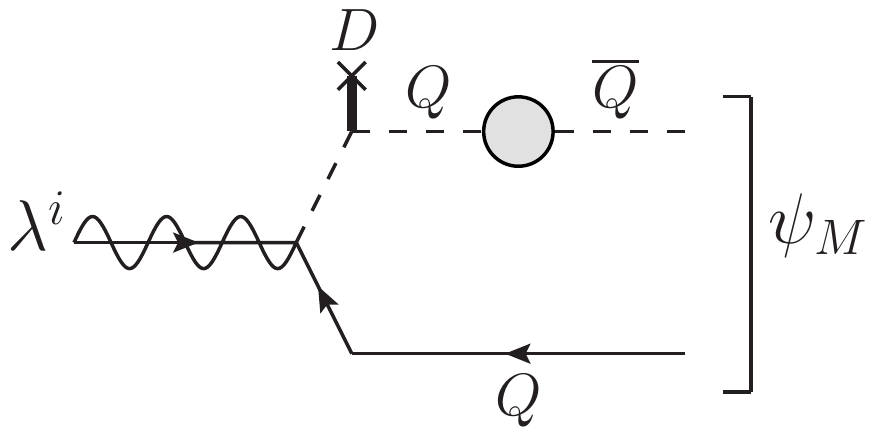}
\linespread{1}
\caption{\small {Schematic constituent diagram of gaugino-mesino coupling generating the Dirac gaugino state in the IR theory. The $D$-term insertion reflects $C$-breaking soft masses for the microscopic squarks ($\tilde m_Q^2 \neq \tilde m_{\overline Q}^2$)} and the blob indicates the strong dynamics underlying the $\langle Q\overline Q\rangle $ condensate. An analogous diagram with $Q\leftrightarrow\overline Q$ also contributes to the Dirac gaugino mass.}
\label{fig:triangle}
\end{center}
\end{figure}

The dynamics underlying the generation of a Dirac gaugino mass is illustrated in Fig.~\ref{fig:triangle}.

We note that the gaugino is odd under a $C$-parity that exchanges $Q$ and $\overline Q$ while the mesino $\psi_M$ is even, so the diagram must involve a breaking of this symmetry. The breaking can enter through a splitting of the squark soft masses\footnote{We can express UV soft masses as projections onto $\Um(1)_A$ and $\Um(1)_B$ directions, such that the mass insertion in Fig.~\ref{fig:triangle} corresponds to a baryon current. We expand on this below.},
which, apart from anomaly-mediated terms, will constitute the only source of SUSY breaking on the UV variables in our setup.

Moreover, the strong dynamics inducing the $\langle Q\overline Q\rangle $ condensate is also responsible for $Q \leftrightarrow  \overline Q$ exchange in the diagram, allowing the external state on the right-hand side to be identified with the mesino $\psi_M$. 


Before proceeding, we note two main advantages we stand to realize in this setup:
\begin{itemize}
\item The desired symmetry breaking vacuum $\langle M\,\rangle \propto {\bf 1}$ in a supersymmetric theory can be a natural consequence of the strong dynamics, rather than being enforced through dedicated model-building;
\item The theory above the strong coupling scale can be governed by a smaller number of degrees of freedom, allowing for  models where SM gauge couplings remain weak below the GUT scale and  the possibility of perturbative unification.
\end{itemize}

\subsection{Realization in $F=N$ SUSY QCD}
\label{sec:SQCD}

The Goldstone gaugino mechanism emerges quite simply in an $\SU(N)$ gauge theory with $F=N$ flavors of quarks $Q$, $\overline Q$, transforming as an $N$ and $\overline N$ of $\SU(N)$, respectively. The theory is described by a manifold of supersymmetric vacua parameterized by meson ($M^i_j$) and baryon ($B, \overline B$) moduli, with certain combinations of these corresponding to physical states at low energy.  The moduli space is deformed at the quantum level, according to the constraint equation
\beq
\label{eq:FNconstraint}
\det M -B\overline B = \Lambda^{2N},
\eeq 
which signals the spontaneous breaking of the $\SU(F)_L \times \SU(F)_R \times \Um(1)_B \times \Um(1)_R$ global symmetry.  Thus the Goldstone superfield providing the  right-handed gaugino can emerge naturally as a composite state of such a theory expanded around a vacuum $ M^i_j  = \Lambda^{2} \delta^i_j$, $B\overline B = 0$ \footnote{As we shall see, it is natural for one of the baryons to acquire a VEV of $\mathcal{O}(\Lambda)$, but since the other baryon will have a large soft mass, it is energetically favorable to satisfy the constraint (\ref{eq:FNconstraint}) with a meson VEV.}  once the appropriate flavor symmetries are weakly gauged.

We therefore assume a vacuum where $M\propto {\bf 1}_N$, in such a way that $N^2 -1$ meson components (the traceless component of $M$) are identified as Goldstones.
We parametrize these fields as
\beq
\label{eq:Mparam}
M = \Lambda\,(\Lambda + \sigma) \, e^{\Pi^i T^i/f} \: , \quad
\Pi^i|_{\theta=\overline\theta=0}=s^i+i\pi^i\: , \quad f \sim \Lambda,
\eeq
with $T^i$ denoting the generators of $\SU(F)_V$ under which $\Pi^i$ transforms as an adjoint\footnote{Due to the incalculability of the Kh{\" a}ler potential away from the origin of the moduli space, it is not possible to determine the numerical ratio between $f$ and $\Lambda$. A canonical Kh{\" a}ler potential of the form $\Delta K = \Tr (M^\dagger M)/\Lambda^2$ would imply $f=\Lambda$ and a canonical normalization of $\sigma$ such that $\sigma\rightarrow \sigma/\sqrt{N}$ in Eq.~(\ref{eq:Mparam}).}.
The constraint (\ref{eq:FNconstraint}) then imposes to leading order
\beq
\label{eq:FNsigma}
\sigma = \frac{1}{N \Lambda^{2N-1}}  B \overline B,
\eeq
removing the singlet meson superfield from the theory.

The Classic supersoft operator in this theory arises as a WZW term once we gauge the diagonal flavor group, $\SU(F)_V$, and identify it with the SM gauge group or one of its simple factors.  Specifically the UV theory possesses an $\SU(F)_A$-$\SU(F)_V$-$\Um(1)_B$ mixed anomaly\footnote{We use this to denote the combination of $\SU(F)_{L,R}^2 \times \Um(1)_B$ anomalies containing the current to which the Goldstones couple.} which is not realized with the IR fermionic degrees of freedom. Instead the anomaly is realized on the Goldstones via a supersymmetric WZW term
\be
\label{eq:FNWZW}
W_\text{WZW} \supset\, \frac{gN}{32\pi^2 f}  \, W_B^\alpha \,W_\alpha^i\,\Pi^i
\ee
with field strengths $W_B^\alpha$ for $\Um(1)_B$ and $W^\alpha$  for $\SU(F)_V$.   This form of the WZW term will be derived in the following subsection starting from SQCD with $F=N+1$ flavors, where all anomalies are saturated by loops of the IR fermions. We will then generate the WZW term and recover the $F=N$ limit by integrating out the flavored baryons of the theory.

Eq.~(\ref{eq:FNWZW}) implies that the $D$-term encoding SUSY breaking in the IR spectrum can be traced to a splitting of the microscopic squark soft masses.  To understand the SUSY-breaking spurion in the IR, we recall that K{\" a}hler terms
$\Delta K = Z Q^\dagger Q$
admit a background axial $\Um(1)$ gauge symmetry
\beq
Q \to e^{-\Omega} Q, \quad 
\overline Q \to e^{-\Omega^\dagger} \overline Q, \quad 
Z \to e^{\Omega+\Omega^\dagger} Z,
\eeq
with $\Omega$ a chiral superfield and $\ln Z$ transforming as the axial $\Um(1)$ gauge field
\beq
\ln Z \to \ln Z + \Omega + \Omega^\dagger \, .
\eeq
This allows the construction of an invariant `field strength' for the vector superfield $\ln Z$; cf. \cite{ArkaniHamed:1998kj} for details.
Allowing for $\tilde m_Q^2 \neq \tilde m_{\overline Q}^2$ via distinct wave function factors for the fields, we thus have
\beq
\begin{split}
W_B^\alpha &=\, -\frac{1}{4}\overline D^2 D^\alpha \ln \left( \frac{Z_Q}{Z_{\overline Q}} \right)  \\
&\supset \, 2\,  \tilde m_{\mathcal V}^2 \, \theta^\alpha ,
\end{split}
\eeq
where we use the notation 
\beq
\label{eq:AVsoft}
\tilde m_{\mathcal A}^2 \equiv \frac{\tilde m_Q^2 + \tilde m_{\overline Q}^2}{2}~~,~~
\tilde m_{\mathcal V}^2 \equiv \frac{\tilde m_Q^2 - \tilde m_{\overline Q}^2}{2} 
\eeq  
for soft masses projected along the axial and vector $\Um(1)$ directions respectively.
We thereby recover the Dirac gaugino mass generated by the Classic supersoft operator (\ref{eq:FNWZW}) as
\beq
m_D = \frac{gN}{16\pi^2} \frac{\tilde m_{\mathcal V}^2}{f}.
\eeq

Soft masses for the microscopic SQCD squarks are expected also to feed into the real scalar components of the Goldstone superfields.  If we assume that the lowest order K\"ahler term for the mesons is of the form
\beq
\label{eq:Knaive}
\Delta K \propto \frac{Z_Q Z_{\bar Q}}{\Lambda_I^2} \Tr  (M^\dagger M)
\eeq
with normalization related to the invariant scale $\Lambda_I = \Lambda^\dagger Z_Q^{F/b} Z_{\bar Q}^{F/b} \Lambda$ of \cite{ArkaniHamed:1998wc}, then expanding in powers of $\theta$ reveals soft mass terms for the scalars, $s$:
\beq
\tilde m_s^2 \propto \tilde m_{\mathcal A}^2.
\eeq
The pseudoscalars, on the other hand, are left massless by (\ref{eq:Knaive}); this is exactly as required by simple symmetry arguments, since the SQCD squark masses do not break the global symmetry, meaning that the masses of the $\pi$ fields remain protected by Goldstone's theorem.

One potential issue that needs to be addressed is the generation of a tachynic soft mass for one of the baryons in the presence of a $U(1)_B$ $D$-term. The easiest way to see that perhaps is to introduce a $\Um (1)_B$ Fayet-Iliopoulos term, and to gauge $\Um (1)_B$ so that the auxiliary field $D$ couples to the scalar baryons in the K\"ahler potential in the usual way. After integrating out the auxiliary field $D$, a tachyonic mass is induced for one of the baryons. This need not be a problem, however, since the SQCD baryons do not carry SM charges. In this case, we expect a K\"ahler potential (neglecting interactions with other fields) containing a general polynomial in $B$ and $B^*$, with the leading term being $-\tilde m_{\mathcal V}^2 B B^*$ and with higher terms suppressed by the scale $\Lambda$. For this scenario to be viable, this baryonic direction cannot be a runaway, but instead the polynomial must have the (very reasonable) property that it has a minimum {\em somewhere} with characteristic value $\Lambda$. Alternatively, these baryons could have mass from a higher dimension operator $\mu^4\, Q^3 \overline Q^3/M_*^3$, with $M_* > \Lambda$. For this to work, however, both $M_*$ and $\Lambda$ should be very high scales to allow the mass $\mu^4/M_*^3$ to be larger than $\tilde m_{\mathcal V}$ and compensate for the tachyonic soft mass. Regardless, this tachyon is a far less serious concern in that it does not apply to any of the fields which must be present in the low energy theory, and a baryon VEV does not cause any fundamental problems so long as it does not run away.

Lastly, we comment on the matching of additional anomalies between the UV and IR theories. In ordinary SQCD with \emph{ungauged} flavored symmetries, there is, for example, a $U(1)_R\,\SU(F)^2$ anomaly matched in the IR by the mesino $\psi_M$. One might naively worry that this anomaly is not matched in the deep IR once the mesino marries the SM gaugino and is integrated out. That is not the case, however. Differently from SQCD with ungauged flavor symmetries, in our scenario the gauging of some flavor symmetries introduces extra fermionic degrees of freedom to the theory, namely the SM gauginos. These extra fermions contribute so as to cancel the $U(1)_R\,\SU(F)^2$ anomaly, hence preserving the t'Hooft anomaly matching conditions. The same reasoning applies to all the other anomalies realized by the composite fermions in our scenario with gauged flavor symmetries.

\subsection{Connecting with $F \neq N$}
\label{sec:Np1toN}
The structure of the low energy theory with $F=N$ can be illuminated by starting in a theory with $F=N+1$ flavors --- whose details we first briefly recap --- and adding a supersymmetric mass for one flavor.  

In SQCD with $F=N+1$, the origin remains on the quantum moduli space  (that is, the theory $s$-confines), and all of the mesons and baryons correspond to light physical states. We will denote these as $\cal M$ and $\cal B, \overline{\cal B}$ respectively.  As the baryons are decoupled via the theory's superpotential $W = \cal B\, \cal M\, \cal{\overline B}\;$ by moving away from the origin, any anomalies matched by these baryons must translate into WZW terms in the surviving $F=N$ theory.  This non-decoupling effect of the baryons is clearly seen by performing a standard perturbative loop calculation, and mirrors the role of `constituent quarks' in ordinary QCD \cite{Manohar:1983md}.

In the $F=N+1$ theory the meson directions develop VEVs proportional to supersymmetric $Q\overline Q$ mass terms.  With $\Lambda_{N+1}$ indicating the strong coupling scale in this theory, the vacuum is described by
\beq
\label{eq:Np1vacuum}
\langle (\mathcal M^{-1})^i_j \det \mathcal M \rangle = m^i_j \Lambda_{N+1}^{2N-1}
\eeq
indicating that fluctuations of the composites do not lie on the moduli constraint surface.  Classical relationships between the composites therefore arise as equations of motion, which can be appropriately arranged with
\beq
\label{FN1W}
W = \frac{1}{\Lambda_{N+1}^{2N-1}} \left( \mathcal B \mathcal M \overline{\mathcal B} - \det \mathcal M \right).
\eeq
Baryons thus become massive as $\langle \mathcal M \rangle \neq 0$.

To recover the $F=N$ case, we introduce a mass for the $F$th flavor, 
\beq
\label{eq:Np1toN}
\Delta W = m Q^F \overline Q_F = m \mathcal M^F_F,
\eeq 
and decompose the composite fields as
\beq
\label{eq:Np1decomposition}
\mathcal M = \begin{pmatrix} M & X \cr \overline X  & \mathcal M^F_F \end{pmatrix}, 
\quad \mathcal B = \begin{pmatrix} Y \cr B \end{pmatrix}^T, 
\quad \overline{\mathcal B} = \begin{pmatrix} \overline Y \cr \overline B \end{pmatrix},
\eeq
where $M$, $X$, $\overline X$, $Y$, $\overline Y$ are in the $\square\times\overline{\square}$, ${\bold 1}\times\overline{\square}$, $\square\times\bold 1$, $\overline\square\times\bold 1$ and ${\bold 1}\times\square$ representations of $\SU (F\!-\!1)_L \times \SU (F\!-\!1)_R$, respectively, and $\mathcal M^F_F$, $B$, and $\overline{B}$ are singlets.
The vacuum condition (\ref{eq:Np1vacuum}) becomes
\beq
\det M - B \overline B = m \Lambda_{N+1}^{2N-1} = \Lambda_{N}^{2N}
\eeq
where the usual scale matching between the $F=N+1$ and $F=N$ theories provides the second equality. The $Y$, $\overline Y$ and $X$, $\overline X$ fields become massive, and can be integrated out. We shall now discuss the effects of integrating out $Y$, $\overline Y$.


The $Y$, $\overline Y$ baryons mediate mixed $\SU(F)^2\,\Um(1)_B$ anomalies, and once integrated out (taking $m \to \infty$), the $F=N$ limit is recovered and WZW terms arise to realize those anomalies. Indeed we will check appropriate diagrams explicitly. Using the parameterization for $M$ in (\ref{eq:Mparam}), the pertinent superpotential couplings are
\beq
\Delta W = m\,  Y \,e^{\Pi/f} \, \overline Y,
\eeq
where $Y\,,\overline Y$ have been normalized to have canonical dimension, i.e. $Y/\Lambda_{N}^{N-1}\rightarrow Y$ and $\overline Y/\Lambda_{N}^{N-1}\rightarrow \overline Y$ in Eq.~(\ref{FN1W}). At linear order we obtain the following interactions for the scalar components of $\Pi$:
\beq
\label{eq:Ymes}
\Delta \lag = -\frac{m}{f} (1+  s^i+i\pi^i)\,\psi \,T^i\,\psi^c + {\rm h.c.}
\eeq
where $\psi, \psi^c$ are the (left-handed Weyl) fermion components of $Y$ and $\overline Y$.
We can now proceed with familiar component calculations to recover the various pieces of the $F=N$ theory's WZW term, Eq.~(\ref{eq:FNWZW}).

Consider the pseudoscalar $\pi$ for example.  From a three-point  diagram involving $\pi$ and two gauge bosons we extract the $\mathcal O(p^2)$ piece, which is the only term that survives as $m \to \infty$; one fermion propagator therefore contributes proportional to its mass rather than momentum.  Hence the non-decoupling effect of $Y, \overline Y$ on the $\SU(F)_A$-$\SU(F)_V$-$\Um(1)_B$ anomaly gives:
\beq
\label{eq:Aloop}
{\raise-1.2cm \hbox{\includegraphics[height=3.3cm]{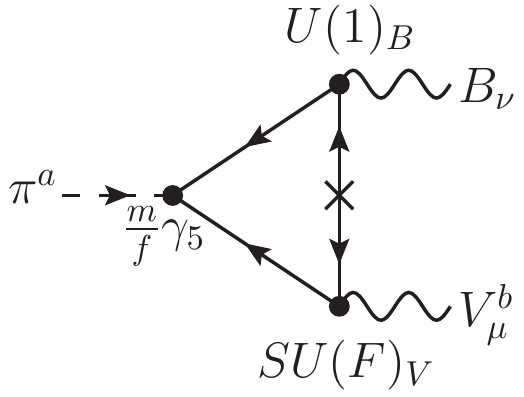}}}  
&+&  \rm permutations  \nonumber \\
 &=& 4N \Tr \left(T^a T^b \right) \Tr \left(\gamma^5 \gamma^\mu \gamma^\nu \gamma^\alpha \gamma^\beta \right)  p^{(1)}_\alpha p^{(2)}_\beta \int \frac{d^4 k}{(2 \pi)^4} \frac{im}{f} \frac{m}{(k^2 -m^2)^3} \nonumber \\
&=& -\frac{N}{4 \pi^2f} \, \delta^{ab} \, \epsilon^{\mu \nu \alpha \beta}  p^{(1)}_\alpha p^{(2)}_\beta \, .
\eeq
Here the cross indicates the mass insertion $m$; $p^{(1,2)}$ are the momenta entering the gauge vertices, and the factor of $N$ comes from the $\Um(1)_B$ charge of the fermions.\footnote{We have taken baryon number to be gauged in this component calculation. We may ultimately turn its gauge coupling off since the $\Um(1)_B$ field strength needs to act only through its $\mathcal O(\theta)$ SUSY breaking spurion.}  This result implies a WZW term \cite{Wess:1971yu, Witten:1983tw} in the low-energy Lagrangian,
\beq
\label{eq:WZlag}
\Delta \lag  = -\frac{gN}{16 \pi^2f} \pi^i \,  F_{\mu \nu}^i  \tilde B^{\mu \nu}\, .
\eeq
This particular derivation parallels that of the $\pi^0 \to \gamma \gamma$ computation with constituent quarks in QCD \cite{Manohar:1983md}.

The WZW term (\ref{eq:WZlag}) can be written as part of a superpotential, and we can confirm with similar manipulations to those above that terms related to it by SUSY are consistently reproduced by taking
\beq
\Delta W =  \frac{gN}{32 \pi^2f} \, W_B^\alpha \,W^i_\alpha\, \Pi^i,
\eeq
and hence we have recovered the Classic supersoft operator (\ref{eq:FNWZW}). 

\section{\label{sec:perturb} A Perturbative Model}


Even though strongly-coupled models offer a natural UV completion for Goldstone Gauginos, perturbative and renormalizable UV completions are also possible. In this section we discuss one such possibility. The basic ingredients discussed in Sec.~\ref{sec:GG} are easily incorporated, such as a messenger sector coupling to bifundamentals of an approximate chiral symmetry containing the SM as a subgroup. To ensure the generation of the Classic supersoft operator, the spontaneous breaking of the global symmetry must be arranged such that the currents associated with the Goldstone fields are anomalous with respect to the SM and the U(1) carrying the $D$-term.

In the model we will discuss, a larger messenger sector will be necessary to avoid destabilizing the symmetry breaking vacuum. The model will consist simply of two replicas of the toy model discussed in Sec.~\ref{sec:GG}:
\be
W_{\text{mess}} = \lambda\; \overline T_1\, \Sigma_1\, T_1 + \lambda\; \overline T_2\, \Sigma_2\, T_2,
\label{eq:pertmess}
\ee
where the equality of couplings is enforced by an exchange symmetry $1\leftrightarrow 2$. We will discuss the symmetry breaking sector shortly. This messenger sector by itself is invariant under an enlarged global symmetry, $\SU(F)_{L1}\times \SU(F)_{R1}\times\Um (1)_{T1}\times\Um (1)_{A1}$ and $\SU(F)_{L2}\times \SU(F)_{R2}\times\Um (1)_{T2}\times\Um (1)_{A2}$, acting independently on each of the two replicas. The $A$ subscript denotes an axial symmetry.  Supposing for the moment that both $\Sigma_{1}$ and $\Sigma_{2}$ would obtain diagonal VEVs, $\langle\Sigma_{1}\rangle=\langle\Sigma_{2}\rangle=f\times\bold{1}$, the global symmetry would be broken to the diagonal subgroups, $\SU(F)_{V1}\times\Um (1)_{T1}$ and $\SU(F)_{V2}\times\Um (1)_{T2}$, and the low energy theory would have $2 F^2$ Goldstones, $\Pi_1$ and $\Pi_2$, such that:
\be
\Sigma_{1,2}=(f+\sigma_{1,2})\,e^{\Pi_{1,2}/f}\qquad,\qquad\Pi_{1,2}|_{\theta,\overline\theta=0}=s_{1,2}+i\,\pi_{1,2}.
\ee
Moreover, there would be two independent mixed anomalies as displayed below.
\vspace{0.4cm}
\begin{figure}[h]
\begin{center}
\includegraphics[height=2.4cm]{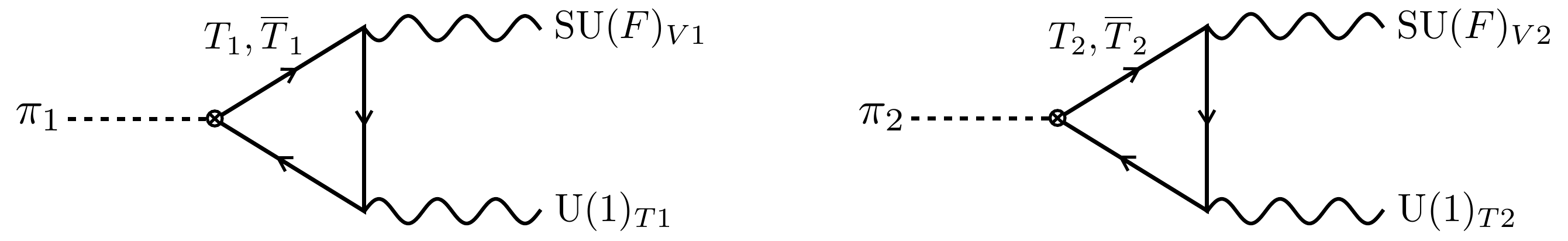}
\end{center}
\end{figure}
\vspace{-0.4cm}

The symmetry breaking sector, however, will not respect the full chiral symmetry of the messenger sector, but only the chiral subgroup $\SU(F)_{L}\times \SU(F)_{R}\times \Um (1)_{A}$ where $L=L_1+L_2$ and $R=R_1+R_2$, under which $\Sigma_1$ and $\Sigma_2$ transform as $(\square,\overline\square,1)$ and $(\overline\square,\square,-1)$, respectively. In reality, therefore, there will only be $F^2$ Goldstone bosons resulting from the spontaneous breaking $\SU(F)_{L}\times \SU(F)_{R}\times \Um (1)_{A}\rightarrow\SU(F)_{V}$, given by the linear combination $\pi = \pi_1 - \pi_2$. The mixed anomaly sourcing the Classic supersoft operator will be $\SU(F)_A$-$\SU(F)_V$-$\Um(1)_T$,\footnote{Due to the doubling of messengers there are in principle two different $\Um(1)_T$ symmetries.  We allow a non-zero D-term for the combination under which the charge of $T_1$ is opposite to the charge of $T_2$} schematically displayed below (where $T=T_1+T_2$ denotes the messenger number):
\vspace{0.4cm}
\begin{figure}[h]
\begin{center}
\includegraphics[height=2.4cm]{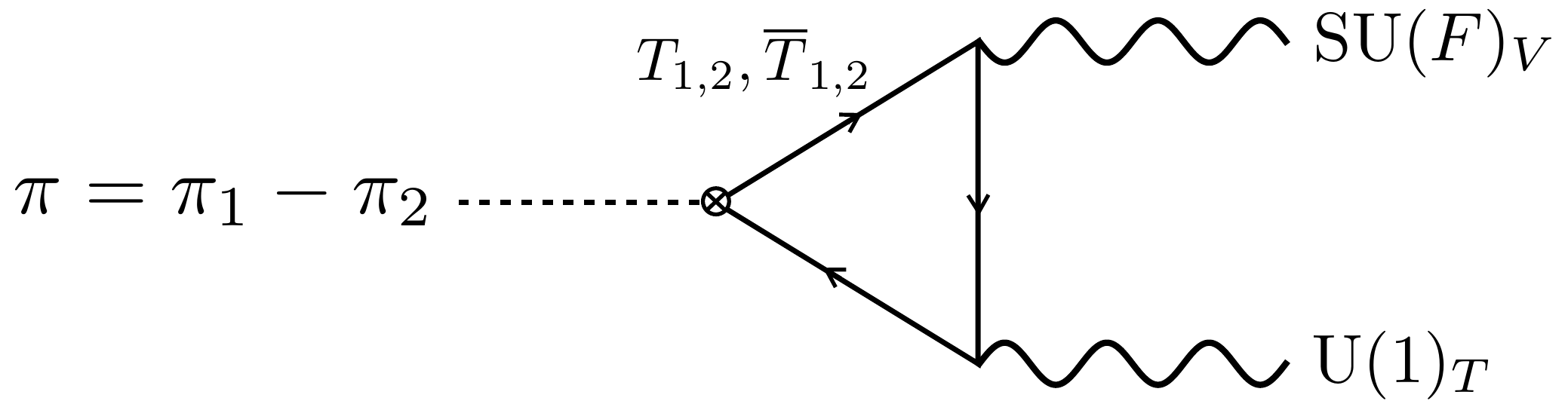}
\end{center}
\end{figure}
\vspace{-0.4cm}

\noindent This mixed anomaly is manifested in the IR by a WZW term,
\be
W_{\text{WZW}} \supset \frac{g}{32\pi^2f}\;W_T^\alpha\, W^i_\alpha\, \Pi^i,
\ee
where $\Pi^i=\Pi^i_1-\Pi^i_2$. As before, we recognize this as the Classic supersoft operator, realizing the Goldstone Gaugino mechanism once it is arranged for $\,\langle W_T^\alpha\, \rangle= D\,\theta^\alpha$.


We can now finally introduce the spontaneous symmetry breaking sector,
\be
W_{\text{SSB}} = \kappa_1\, S \left( \Tr \big[ \Sigma_1\,\Sigma_2 \big] - v^2 \right) + \kappa_2 \Tr \big[ \Sigma_1\,\Omega\,\Sigma_2 \big] ,
\label{eq:pertModel}
\ee
where, as before, $\Sigma_1$ and $\Sigma_2$ are in the $(\square,\overline\square,1)$ and $(\overline\square,\square,-1)$ representations of $\SU(F)_{L}\times \SU(F)_{R} \times \Um (1)_{A}$, respectively, and we have introduced two new chiral superfields $S$ and $\Omega$ transforming as $(\bold{1},\bold{1},0)$ and $(\bold{1},\bold{Ad},0)$, respectively.

The first term in (\ref{eq:pertModel}) induces a VEV to the trace components of $\Sigma_1\,,\,\Sigma_2$, such that $\langle\Sigma_{1}\rangle= \alpha\, v/\sqrt{F} \times \bold{1}$ and  $\langle\Sigma_{2}\rangle= \alpha^{-1} \,v/\sqrt{F} \times \bold{1}$. The ratio of these VEVs, parameterized by $\alpha$, need not be one; it represents a flat direction of the potential that in principle will be lifted by SUSY breaking. If $\alpha \neq 1$ the low energy Goldstones would be given by a different admixture of $\Pi_1$, $\Pi_2$, but otherwise our previous discussion of the low energy theory remains unchanged. 

Finally, the second term in the superpotential (\ref{eq:pertModel}) is present to give masses to the remaining degrees of freedom upon spontaneous breaking of the global symmetry. The components of $\Omega$ pair up with the combination of $\Pi_1$ and $\Pi_2$ orthogonal to the Goldstone $\Pi$, and the trace components of $\Sigma_1$ and $\Sigma_2$ orthogonal to the axial Goldstone pair up with the singlet $S$ to obtain a Dirac mass.  Thus there are $(F^2-1)$ Goldstones in the adjoint of the unbroken $\SU(F)_{V}$ symmetry, which may be identified with a SM gauge group, and a singlet Goldstone associated with the spontaneous breaking of the $\Um (1)_{A}$ axial symmetry.  Coupling of either of $\Sigma_1$ or $\Sigma_2$ to a pair of messenger fields leads to a tadpole term for this singlet which picks a vacuum $\alpha \neq 1$, however equal couplings of each field to its own pair of messengers in \Eq{eq:pertmess} avoids this issue as each tadpole term cancels and the calculation is consistent with remaining at the minimum of the scalar potential.

We have shown with this example that it is possible to construct simple renormalizable and perturbative models of Goldstone Gauginos. In general, all the arguments we have presented carry forward when we parametrize the Goldstone fields as exponentials. At the same time, we should be able to see the absence of the lemon-twist by studying the linearized model and including all of the heavy fields as well. In Appx.~\ref{appx:tadpole}, we revisit the non-renormalizable toy model in (\ref{GGtoy}) in the linear representation of the Goldstone fields; we demonstrate that lemon-twist is not generated due to a cancelation of scalar soft masses by a tadpole induced shift in the trace component of $\Sigma$.

\section{\label{sec:conc} Implications and Discussion}

\subsection{Fine-Tuning and Naturalness}
\label{conc1}

With the tachyon problem removed, we are able to ask the question of the naturalness of the theory. While elaborate studies of fine-tuning of the electroweak scale are beyond our scope, a brief discussion of naturalness in terms of radiative corrections to the Higgs mass is warranted.

In these theories, heavy stops are insufficient to explain the Higgs mass due to the dramatic reduction in the $D$-term quartics \cite{Fox:2002bu}, and so something along the lines of the NMSSM must be invoked. On the other hand, the presence of additional matter allows for larger values of the NMSSM quartic that remain perturbative in the UV \cite{Espinosa:1998re,Barbieri:2007tu}, so there is a natural synergy within the setup. Even absent $D$-term quartics, we estimate that for an NMSSM strongly coupled at the GUT scale, we can have a Higgs at 125 GeV for $1.5 \lesssim \tan \beta \lesssim 2$ with sub TeV stops.

The Higgs receives radiative corrections from the winos and binos at one loop, and from the gluinos at two loops (via the usual stop coupling). These are \cite{Fox:2002bu}:
\bea
\left. \delta \tilde m_h^2 \right|_{\rm 1-loop} &=& \frac{2 \alpha_2 \ln(4)}{3 \pi} m_{D2}^2,\\
\left. \delta \tilde m_h^2 \right|_{\rm 2-loop} &=& \frac{3}{8 \pi^2}y_t^2 m_{\tilde t}^2 \ln(m_{D3}/\tilde m_{\tilde t}),\\
 &=& \frac{\alpha_3 y_t^2 \ln(4)}{2 \pi^3} \ln(3 \pi/4 \alpha_3 \ln(4)) m_{D3}^2, \nonumber \\
\tilde m_{\tilde t}^2 \simeq \tilde m_{\tilde q}^2 &=& \frac{4 \alpha_3 \ln(4)}{3 \pi} m_{D3}^2.
\eea
Here $m_{D2}$ and $m_{D3}$ are the masses of the wino and gluino, respectively, and we neglect the contribution from the bino.

\begin{figure}[htb]
\begin{center}
\includegraphics[width=8cm]{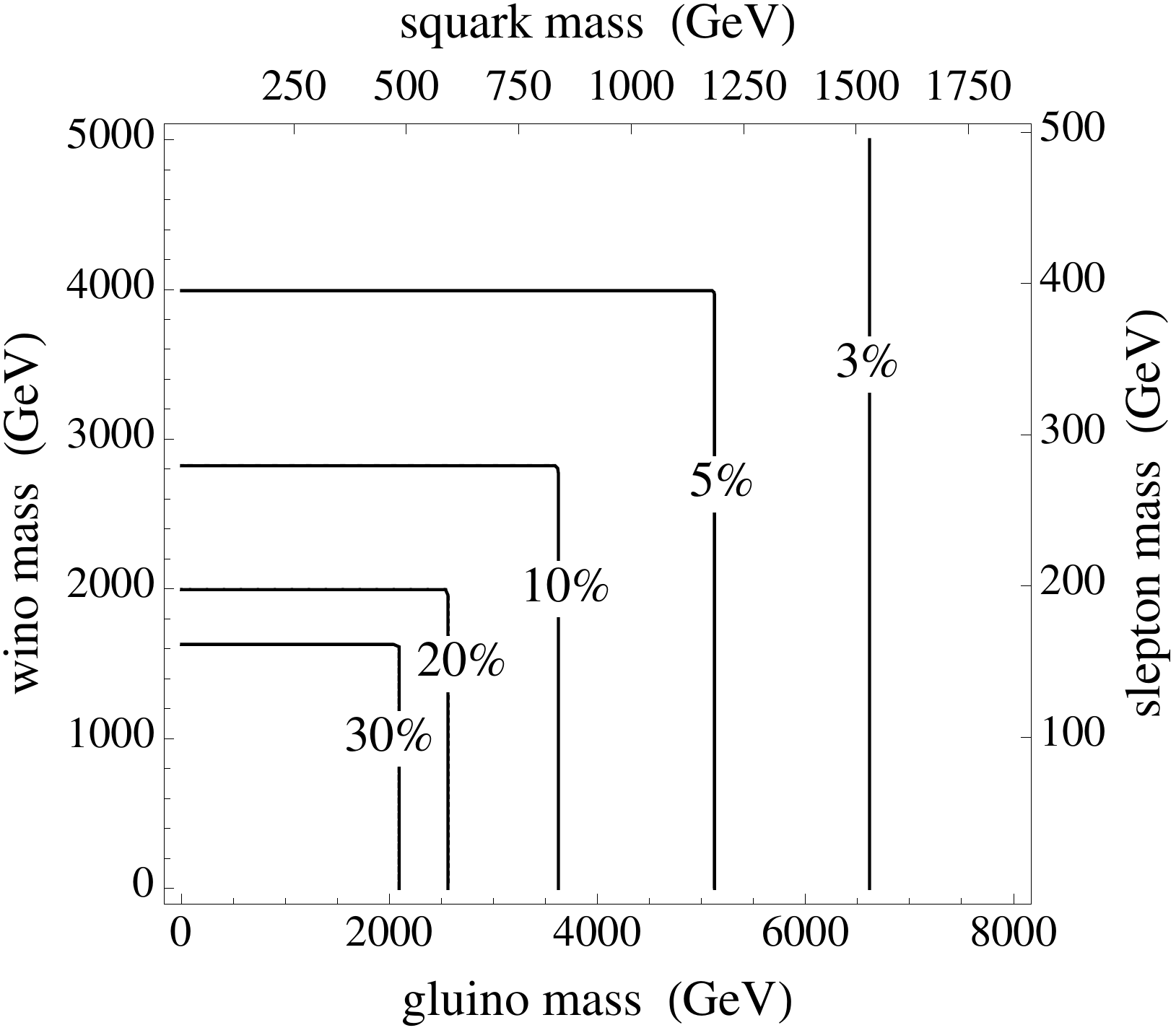}
\linespread{1}
\caption{\small Contours of EW scale tuning (shown as percentage), determined by the ratio of the largest radiative correction of the Higgs mass to its observed value of $(125 \gev)^2$. Radiative contributions to squark and slepton masses from the associated gaugino mass are also shown.}
\label{fig:tuning}
\end{center}
\end{figure}

We can make a rough assessment of the tuning of the theory by evaluating the ratio of these extra soft mass contributions to the Higgs potential relative with the observed value $m_h^2 = (125 \GeV)^2$. That is, we define the tuning to be $\delta \tilde m_h^2/(m_h^2/2)$, and display tuning contours in Figure \ref{fig:tuning} as a function of gaugino and scalar masses.

We see that for a gluino mass in the 2-3 TeV range, the tuning from the gluino contribution is 30\%-15\%, with squarks in the 450-700 GeV range.\footnote{Note that in the MSSM such masses would be at strong tension with existing searches. However, with heavy Dirac gauginos,``doubly invisible'' decay topologies for squarks can be naturally realized \cite{Alves:2013wra}, under which these masses are viable.} This marks a significant improvement over the MSSM with a comparable gluino mass.

\subsection{UV Physics and Unification}
\label{conc2}

Perturbative unification remains a challenge for Dirac gaugino models. A full {\bf 24} of SU(5) contains enough matter that the gauge couplings hit a Landau pole before the GUT scale \cite{Moroi:1993zj, Brahmachari:1993zi}. 
In the context of composite models of Goldstone Gauginos, we have interesting possibilities, however. It is possible that the low energy theory contains SM adjoints built out of confining fundamentals. For instance, one can consider a gauge group $\SU(3)_a \times \SU(3)_b \times \SU(3)_{\rm SM} \times \SU(2)_{\rm SM} \times U(1)_{\rm SM}$, with an $a \leftrightarrow b$ symmetry of couplings imposed at the GUT scale. With fields $d_a = ({\bf 3}, {\bf 1}, {\bf 3}, {\bf 1}, {\bf 1/3})$ and $\ell_b = ({\bf 1}, {\bf 3}, {\bf 1}, {\bf 2}, {\bf 1/2})$ (and vector-like partners), the SM $\beta$ functions are shifted by the equivalent of three flavors. Such a model could arise out of an orbifold GUT, for instance. 

When $\SU(3)_a$ and $\SU(3)_b$ confine, the theory produces composite adjoints of the SM gauge group, and in addition two sets of baryons with hypercharges $\pm 1$, and doublets with hypercharge $\pm 1/2$. Together, this maintains a unified shift to the SM $\beta$-functions. Indeed, this is quite analogous to the (charged) matter present in the adjoint in trinification, but the high energy theory is arising from a set of split SU(5) multiplets. It is difficult to employ this model, however, as the tachyonic baryons would have hypercharge. This could be solved with explicit masses for the microscopic quarks of the strongly coupled sector, but then supersoftness would only be retained if $\Lambda \sim m_{\rm SUSY}$.

As a first alternative, we can look to models where the fields $d_a$ carry no hypercharge; these models do not generically unify, so we will discuss a unifying case subsequently.  Considering the $\SU(3)_b$ gauge group to be associated with its own $\SU(3)_{bL} \times \SU(3)_{bR}$ chiral symmetry, the idea is to identify hypercharge with diagonal subgroup's abelian generator $\lambda^8 \propto {\rm diag}(1,\, 1,\, -2)$;  tracelessness then ensures that all baryons will be completely neutral under the SM gauge group.  Specifically, let us take
\beq
Y= \sqrt 3 \, c\, \lambda^8 =  {\rm diag}(c/2, \, c/2, \, -c).
\eeq
Relative to the wino, the bino's mass is thus scaled by an amount
\beq
\label{eq:Yoverlap}
\Tr \, ( Y \lambda^8)/T_\square = \sqrt 3 \, c\, ,
\eeq
where the index $T_\square$ appears due to our normalization $\Tr \, (\lambda^i \lambda^j) = \delta^{ij}/2$.   The confining group, $\SU(3)_b$, produces an octet of Goldstone bosons, four of which are accounted for by bachelor doublets of hypercharge $y_D = 3 c/2$, and from which we identify the interesting cases $c=1$ and $c = 1/3$ which result in doubly- and singly-charged bachelors, respectively.  Assuming an $a\leftrightarrow b$ symmetry as above, such that the SUSY breaking $D$-terms of each sector are equal, we conclude the following ratios for SM gaugino masses for cases indicated by the EM charge of their bachelor doublets:
\beq
\begin{split}
\mbox{bino : wino : gluino} \; \big|_{Q_D = \pm 1} \, &=\ 1\, : \, 3.2 \, : \, 5.2 \\
\left. \mbox{bino : wino : gluino} \; \right|_{Q_D = \pm 2} \, &=\ 1\, : \, 1.05 \, : \, 1.75
\end{split}
\eeq

In the interest of maintaining perturbative unification, we consider also the case of trinification \cite{Rizov:1981dp, trin2}, where the SM is embedded into $SU(3)^3$.  Here we would require three copies of the gauge$\times$global structure $\SU(3)_i \times \SU(3)_{i,L}\times \SU(3)_{i,R}$, and identify hypercharge as
\beq
Y = -\frac{1}{\sqrt 3} (\lambda^8_{2V} + \lambda^8_{3V}) - \lambda^3_{3V}
\eeq
with subscripts $(i,V)$ denoting generators of the diagonal subgroup of $\SU(3)_{i,L} \times \SU(3)_{i,R}$.  In this case there are three singlet mesons (originating from the $\lambda_{2V,3V}^8$ and $\lambda_{3V}^3$ adjoint directions) that couple to the bino.  Taking the pertinent traces as in Eq.~(\ref{eq:Yoverlap}), we find a mass matrix with two degenerate eigenvalues indicating the two components of the Dirac bino, along with two zero eigenvalues corresponding to singlet bachelors.  Specifically we find eigenvalues with coefficient $\sqrt{5/3}$, indicating a universal gaugino spectrum up to differences in gauge couplings.  With this, the predicted mass ratios become
\beq
\mbox{bino : wino : gluino}  \ = \ 1\, : \, 1.5 \, : \, 2.4 \, .
\eeq

We see that there is sizable variability for the bino mass relative to the other SM gauginos.  The connection between anomalies and the Dirac mass can be quite enlightening here. While the gluino and wino masses must - at some level - arise from a breaking of a flavor symmetry where the diagonal component is identified with the SM gauge group, the same does not hold for the bino. Indeed, {\em any} global symmetry $\Um(1)_X$ that is spontaneously broken and contains a B-Y-X anomaly will yield a Dirac mass mixing $\pi_X$ and the bino. For instance, one could supplement the perturbative models we have described with additional messengers, with a spontaneously generated mass which is flavor diagonal from the outset. The axion of that spontaneous breaking would then marry the Bino. Ultimately, there is one linear combination of fields that will marry the bino, and the important point is that its mass might be quite different from that expected from naive ideas of unification. For instance, if one considers a high scale model, with a large number of Goldstone fields, one could end up with a surprisingly massive bino, while still maintaining an expected GUT relation between the gluino and wino.

Of course, this all assumes we insist on maintaining some embedding within a simple group. We can also consider situations where near the Planck scale we have the classic supersoft operators generated for gluino, bino and wino, but with no bachelor fields and only adjoints. Alternatively, it may be that only the gluino has an adjoint partner, such as in \cite{Hall:1990hq,Nelson:2002ca},  with slightly smaller Majorana masses for the gauginos, and soft masses for the scalars as in the MSSM. Such a possibility could arise with the Dirac gluino mass generated either from high scale physics or from an intermediate $SU(3)$ SUSY QCD theory as we have described. In these cases, the unification of couplings in the MSSM would be an accident. Any of these possibilities (trinification, embedding in SU(3), gluino only, bachelor-free) are reasonable boundary conditions to consider for Goldstone Gaugino models, as are no doubt many others.

Finally, we note that at least {\em some} GUT scenarios predict the existence of extra doublets with hypercharge $\pm 1/2$ as part of complete GUT multiplets (either bachelors in $SU(3)^3$ or the $3$-$3$-$3$-$2$-$1$ model we have described above). If present, this allows the possibility of a ``Sister Higgs'' NMSSM \cite{Alves:2012fx}\footnote{Similar fields, called ``auxiliary Higgses'' in the setup of induced EWSB \cite{Galloway:2013dma}, may also play a role in this area of model building.}, which produces a Type-I 2HDM model at low energies, which is far less constrained than the Type-II at low $\tan \beta$.

\subsection{Basic Phenomenology}
\label{conc3}
Many BSM searches are sensitive to models with Dirac gauginos, but the potential suppression of the gaugino signal, and the presence of new scalar states can change the focus of these searches.
In particular, colored sgluons $s_c$, $\pi_c$ are likely to be a major discovery channel at hadron colliders.  In all of the models discussed the classic supersoft operator (\ref{eq:classic}) is generated at one loop.  This leads to the desired Dirac gaugino masses ($m_D$) and also soft masses for the scalar sgluons $s_c$, $m_{s_c} =2 m_D$ \cite{Fox:2002bu}.  But since the lemon-twist operator (\ref{eq:lemontwist}) is not generated, the pseudoscalar Goldstone sgluons $\pi_c$ remain massless at that order. The usual supersoft mechanism which generates squark masses similarly leads to Goldstone sgluon squared-masses $m_{\pi_c}^2 =3\, \alpha_3 m_D^2 \ln(4)/ \pi$ \cite{Fox:2002bu}.  This leads to the generic expectation that the Goldstone sgluon masses are comparable to the squark masses ($m_{\pi_c} = 1.5 \times m_{\tilde q}$), and are a factor $\sim 3$ suppressed relative to the Dirac gauginos.

The collider phenomenology of sgluons at the LHC has been considered extensively \cite{Choi:2008ub,Plehn:2008ae,Choi:2009jc,Idilbi:2009cc,Choi:2009ue,Idilbi:2010rs,Martynov:2010zz,Han:2010rf,Schumann:2011ji,Aad:2011yh,GoncalvesNetto:2012nt,Kotlarski:2011zz,ATLAS:2012ds,ATLAS:2012nna,Calvet:2012rk,TheATLAScollaboration:2013jha,PD,Chen:2014haa,Renaud:2012pua,Valery:2014pca,Beck:2015cga}.  We will focus on the lightest colored sparticles, which are the Goldstone sgluons and the squarks.  The Goldstone sgluons are neutral under any R-symmetry and couple to gluons at tree-level through standard QCD interactions and pairs of squarks at one-loop. 
They also couple at one-loop to quark-anti-quark pairs with coupling proportional to the quark masses.

At hadron colliders the Goldstone sgluons may be pair produced at tree-level from gluon initiated processes, $gg\to \pi_c\pi_c$, (dominant) or singly produced at one-loop from gluon and quark initial states, $gg,q \overline{q} \to \pi_c$ (subdominant).  The pair production mode is the most promising for collider searches \cite{Plehn:2008ae}.  If the masses for sgluons and squarks are dominated by the radiative contributions we have discussed, the pseudscalar sgluons (psgluons) are kinematically forbidden from decaying to squarks (i.e., $m_{\pi_c} = 1.5 \times m_{\tilde q}$).  They dominantly decay directly to $q \overline{q}$ pairs. Unlike sgluons, psgluons have no trilinear coupling to squarks and their decays arise through gluino loops. The widths are proportional to the masses of the final state quarks, thus Goldstone psgluon pair production may result in top-rich final states.  In particular, the same-sign top final state, $\pi_c\pi_c \to t t j j$, is promising \cite{Plehn:2008ae}.  A number of searches for dijet decays of sgluons have been performed \cite{ATLAS:2012ds} and the (model-dependent) limit on sgluon masses could be as high as $\sim750$ GeV with $20 \text{ fb}^{-1}$ of data at $\sqrt{s}=8$ TeV for decays involving top quarks \cite{Beck:2015cga}.  Taking the upper range of this limit would then imply $m_{D3} \gtrsim 2 \tev$, for which the tuning of the electroweak scale is still quite mild. Searches at LHC14 should improve this limit.

If there are modifications to the radiative mass contributions discussed in this work Goldstone psgluons can decay to pairs of squarks.
In this case the relevant final state involves four squarks which may then decay to neutral LSP states via $R$-parity preserving channels or potentially to SM final states via small R-parity violating interactions.  Any search for these decays is heavily dependent on the spectrum of electroweak states and on the magnitude of various couplings, thus it is model-dependent.

It is not the goal of this work to perform a dedicated collider study.  Nonetheless, it is clear that Goldstone Gaugino models lead to interesting collider signatures which are discoverable at the LHC.

\section{Conclusions}
\label{concl}

The absence of any sign of low scale SUSY pushes us to more tuned models or models that qualitatively differ from the MSSM. Models with Dirac gauginos provide an intriguing possibility of a qualitatively different BSM scenario. In particular, the radiative corrections yield different spectra, and potentially improved naturalness of the electroweak scale compared with the MSSM.

Unfortunately, models of Dirac gauginos generally have tachyons, and previously considered means to remove them have spoiled many of the appealing features of the models. The recently proposed Goldstone Gaugino (GoGa) mechanism provides a simple solution for the absence of charged tachyons in the theory. This allows us to both engage in producing simple models of GoGas, but also to study implications of high scale models where the ``lemon twist'' operator is absent as a boundary condition. 

This simple fact - that the tachyon is absent with Goldstone Gauginos, together with the simple models that realize this - puts Dirac gaugino models on a firmer footing, and helps make them once again a viable candidate for a natural model of the weak scale.

\begin{acknowledgments}

We thank Nathaniel Craig, Sergei Dubovsky, Claudia Frugiuele, Brian Henning, Anson Hook, Patrick Fox, Adam Martin, John Terning, and Jay Wacker for useful discussions. D.A. is supported by NSF-PHY-0969510 (the LHC Theory Initiative). N.W. and D.A. are supported  by the NSF under grants PHY-0947827 and PHY-1316753.  J.G. is supported by the James Arthur Postdoctoral Fellowship at NYU, and thanks the Aspen Center for Physics (National Science Foundation Grant No. PHYS-1066293) for hospitality during this work.

\end{acknowledgments}

\appendix

\renewcommand\theequation{\thesection.\arabic{equation}}

\section{Absence of LTS -- A Diagramatic Proof}
\label{sec:AToyModel}



The absence of lemon-twist in the GoGa setup can be proven in a simple diagrammatic way, which we present here.  The essential point is to show that the interactions between messengers and adjoints, which allow the construction of pseudoscalar mass terms in the conventional Dirac gaugino case, are precisely cancelled once we account for higher order terms in the pion expansion of the superpotential.

The lemon-twist operator comes from a combination of loops constructed from either trilinear or quartic couplings: with the Yukawa coupling and messenger mass terms defined to be real, the mass of $a_R$ receives corrections from loops with trilinear and quartic terms while $a_I$'s mass is generated only via the latter.  The shift symmetry of the Goldstone model then must produce a cancellation of the quartic interactions involving $a_I$, and this is instructive to view diagrammatically.  Each of the above models' messenger auxiliary fields $F_{T, \overline T}$, which we include in diagrams in order to explicitly illustrate differences between the superpotentials, generate interactions such as 
\beq
\label{eq:QMass1}
{\raise-0.9cm \hbox{\includegraphics[height=2.2cm]{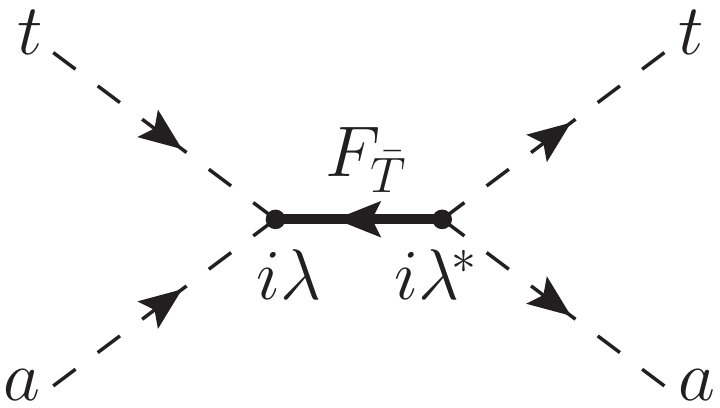}}}  
\quad \Rightarrow \quad \Delta \lag = -|\lambda^2 |\,  t^*t \, a^*a \, ,
\eeq
with an analogous term involving $\overline t$ from taking $T\leftrightarrow \overline T$.  
Closing the $t/\overline t$ line generates a mass at order $D^2$ for both the real and imaginary components of $a$,
\beq
\delta m_{a_{R,I}}^2 = +\frac{\lambda^2}{16 \pi^2}\frac{D^2}{\mu^2},
\eeq
where $\mu$ is the supersymmetric $T \overline T$ mass, identified as $\mu \to \lambda f$ in the Goldstone case.  The trilinear coupling of the real component generates an additional one loop graph giving
\beq
\delta m_{a_R}^2 = -\frac{\lambda^2}{8 \pi^2}\frac{D^2}{\mu^2},
\eeq
and thus $\delta m_{a_R}^2<0$ in total, signaling the problematic instability.

In  the Goldstone model, we must account for the new superpotential terms that arise from expanding $\exp (A/f)$ beyond linear order.  At quadratic order in $A$ the crucial term is $\Delta W = \lambda A^2 T\overline T/2f$, generating
\beq
\label{eq:QMass2}
{\raise-0.9cm \hbox{\includegraphics[height=2.2cm]{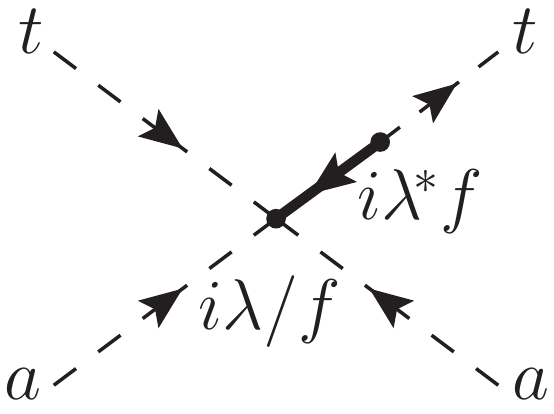}}}  
\quad \Rightarrow \quad \Delta \lag = -\frac{|\lambda|^2}{2} t^* t \, (a^2 + a^{*2}) \, ,
\eeq
again with analogous $\overline t$ terms.  The combination of effects in Eqs.~(\ref{eq:QMass1}, \ref{eq:QMass2}) now gives a vanishing  $t^*t \, a_I^2 $ interaction,  guaranteeing that lemon-twist is not generated since this is the only interaction feeding into the pseudoscalar mass. The real scalar retains cubic and quartic interactions with the messengers, but both are modified in such a way that the two types of loop contributions to $m_{a_R}^2$ precisely cancel. The shift symmetry  prevents loop diagrams from generating masses, so these cancellations come as no surprise. But looking directly at the interactions and their modified versions in the Goldstone model shows where and how these cancellations emerge in an illustrative way.

In contrast, there is no difference between the radiative generation of Dirac gaugino masses in the Goldstone model and in the conventional model. Thus, while we still get the same classic operator, lemon-twist is absent.

\section{Absence of LTS -- Connection to Holomorphic Gauge Coupling}
\label{sec:hgc}

We can explicitly check that the lemon-twist operator is not generated in the GoGa toy model discussed in Sec.\ref{GGtoy} by looking at the renormalization of the $U(1)_T$ holomorphic gauge coupling. In particular, a $\Pi^2$-lemon-twist operator
\be
\label{l-t}
W_\text{LTS} \propto \frac{1}{f^2} \Tr \big[\Pi^2 \big]\,W_T^{\alpha} \, W_{T\alpha} 
\ee
can only be generated if the messengers' masses are corrected when a $\Pi$ background superfield is turned on. Integrating out the messengers corrects the $U(1)_T$ holomorphic gauge coupling below the messenger threshold, and could generate lemon-twist type operators \cite{Csaki:2013fla}.

Recalling that the superpotential in messenger sector is given by:
\be
W=\lambda\; \overline T\, \Sigma\, T,
\ee
and turning on a $\Sigma$ background superfield, we can integrate out $\overline T$, $T$ at $m_T = \lambda \Sigma$. The one-loop holomorphic gauge coupling is then
\be
\label{tau1loop}
\tau_\text{1-loop} = \frac{4\pi i}{g^2(\Lambda)} + \frac{i\,b_L}{\pi}\ln\frac{\mu}{\Lambda} - \frac{i}{\pi}\ln\frac{\text{Det}(\lambda \Sigma)}{\Lambda^{F}},
\ee
where $b_L$ is the $\Um (1)_T$ $\beta$-function coefficient below the $m_T$ threshold.
Analytically continuing $\Sigma$ into superspace, we infer that the following operator is generated at low energies:
\be
\label{1loopOP}
\frac{1}{16 \pi^2} \Tr \log\frac{\lambda\,\Sigma}{\Lambda} ~W^{B \alpha} \, W_\alpha^B 
= \frac{F}{16 \pi^2} \log\frac{\lambda(v+\sigma)}{\Lambda} ~W^{B \alpha} \, W_\alpha^B,
\ee
where we have used $\Sigma=(v+\sigma)\,e^{\Pi^a T^a/f}$ and the fact that the generators $T^a$ are traceless.

The operator (\ref{1loopOP}) is independent of the Goldstone superfields $\Pi$, so the $\Pi^2$-lemon-twist operator (\ref{l-t}) is not generated at one-loop. Moreover, because the one-loop holomorphic gauge coupling is exact, $\Pi^2$-lemon-twist is not generated at any order in perturbation theory.

\section{Absence of LTS -- Tadpole Shift in Linear Representation}
\label{appx:tadpole}

As we have discussed, there are a variety of means to achieve the appropriate flavor symmetry breaking, including a range of both renormalizable and non-renormalizable perturbative models. If we treat the Goldstone Gaugino non-linearly, i.e., as pions $v \,e^{ \Pi/f}$, the protection against the lemon-twist operator is manifest \cite{Alves:2015kia}. However, we can also work with fields expanded linearly around the vacuum, in which case the cancellation is not as obvious. It can be instructive to see how this occurs in those cases.

As we described in Sec. \ref{sec:nonperturb}, a simple model that approximates the QCD dynamics is the superpotential
\be
W= \frac{1}{\Lambda^{N-2}} S\left( \text{Det}\Sigma-v^N \right) .
\label{eq:non-ren}
\ee
As noted previously, this approximates the dynamics of SUSY QCD with an equal number of colors and flavors ($F=N$). In the vacuum, the symmetry breaking structure is $\SU(N)_L \times \SU(N)_R \to \SU(N)_V$ as the $F$-term for $S$ requires a vacuum expectation value $\langle \Sigma \rangle = v\, {\bf 1}_N$.  The $N^2-1$ traceless components of $\Sigma$ become Goldstone superfields of the spontaneously broken symmetry.

In order to accommodate gluinos, a gauge group $N \geq 3$ is required, meaning that these models are not renormalizable and must be UV-completed at a scale $\Lambda$.  Nonetheless, they provide a straightforward implementation of the Goldstone Gaugino scenario and are worth considering further.  We will focus on the case $N=3$, which is appropriate for a model of Dirac gluinos.  If we expand out the superfield $\Sigma$ into its trace component $v+X$ (where $X$ is the perturbation of the trace around the vev) and the traceless component $a$, \emph{i.e.,} $\Sigma = v+X+a$, then in the vacuum the superpotential \Eq{eq:non-ren} becomes
\be
W= \frac{1}{\Lambda} S \left( \text{Det}(a)-\frac{1}{2} (v +X) \Tr[a^2] + (v+X)^3 -v^3 \right).
\label{eq:non-ren2}
\ee
$X$ has paired up with $S$ to obtain a Dirac mass $M_X = 3 v^2/ \Lambda$.  The Goldstone superfield, $a$, remains massless.

If $\Sigma$ is coupled to the SUSY breaking messengers as in \Eq{GGtoy}, then as explained in Sec.\ref{sec:GG}  a $B\mu$-term for $a$ should be generated at one loop. As described in Appx.\ref{sec:hgc}, we can think of the generation of this lemon-twist operator as a correction to the holomorphic gauge coupling. This arises because the messengers, with mass $v$, have their masses corrected in the presence of $a$ backgroung fields. Since $\Tr[a]=0$, this correction enters only at $\mathcal{O}(a^2)$. Moreover, the masses of the messengers depend linearly on $X$ and so the lemon-twist term to leading order is given by
\be
W \supset -\frac{1}{32 \pi^2 v^2} W'^\alpha W_{\alpha}'  \left( -2 v N_f  X +\Tr[a^2] \right),
\ee
where $2 \times N_f=6$ in this case under consideration. We ignore the $\mathcal{O}(X^2)$ term under the assumption that the lemon-twist corrections to the mass of $X$ are small, i.e., $D^2/v \ll 3\, v^2/\Lambda$. This results in $B\mu$-term contributions for $a$ in the scalar potential:
\be
V \supset \frac{1}{2} m_{\rm LT}^2 \Tr[a^2] + \text{h.c.} ~~,
\label{eq:dangBmu}
\ee
where $m_{\rm LT}^2=D^2/16 \pi^2 v^2$ .  In addition to this, a tadpole term for the heavy singlet scalar is also generated
\be
V \supset -3\, v\, m_{\rm LT}^2 X + \text{h.c.}
\ee

When the scalar potential is minimized this leads to an additional contribution to the symmetry breaking VEV,
\be
\langle X \rangle = \frac{3\, v\, m_{\rm LT}^2}{M_X^2} =  \frac{m_{\rm LT}^2 \Lambda^2}{3\, v^3}.
\ee
Inserting this shifted vacuum into the $F$-term equation for $S$, to lowest order in $m_{\rm LT}^2$ we have
\be
\mb{F_S} =\frac{1}{\Lambda} \left( \text{Det}(a)-\frac{1}{2} \left(v+\frac{m_{LT}^2 \Lambda^2}{3\, v^3} \right) \Tr[a^2] +\frac{m_{\rm LT}^2 \Lambda^2}{v} \right)
\ee
Squaring this $F$-term to solve for the scalar potential we find that the dangerous $B\mu$-term of \Eq{eq:dangBmu} is exactly cancelled.
This is not surprising - the symmetries alone dictate this, and in the nonlinear realization it is manifest. Working with linear fields around the origin, however, this is much more involved: in this formulation the absence of $B\mu$ terms occurs through cancellations of the explicit mass terms generated from messenger loops against new $B\mu$ terms generated by a shift in the VEV of a heavy field in the symmetry breaking sector. Such a scenario shows clearly that such cancellations can occur in a very non-trivial fashion when the Goldstone nature of the right-handed gaugino is obscured.

\bibliographystyle{JHEP}
\bibliography{REFS}

\end{document}